# Two-electron bunching in transport through a QD induced by Kondo correlations

O. Zarchin, M. Zaffalon, M. Heiblum, D. Mahalu, and V. Umansky

*Braun Center for Submicron Research, Department of Condensed Matter Physics,*

*Weizmann Institute of Science, Rehovot 76100, Israel*

## ABSTRACT

We report on noise measurements in a quantum dot in the presence of Kondo correlations. Close to the unitary limit, with the conductance reaching $1.8e^2/h$, we observed an average backscattered charge of $e^* \sim 5e/3$, while weakly biasing the quantum dot. This result held to bias voltages up to half the Kondo temperature. Away from the unitary limit, the charge was measured to be $e$ as expected. These results confirm and extend the prediction by E. Sela *et al*. [1], that suggested that two-electron backscattering processes dominate over single-electron backscattering processes near the unitary limit, with an average backscattered charge $e^* \sim 5e/3$.

The Kondo effect is a many-body problem resulting in the formation of a dynamical singlet between a localized spin impurity and the delocalized conduction electrons [2]. The low temperature Hamiltonian of the Kondo problem [3] contains a term involving two-electron correlations, leading to bunching of the scattered electrons. To reveal such correlations, we fabricated a quantum dot (QD), being a confined region in a two-dimensional electron gas, separated from two electron reservoirs by two tunnel barriers. Owing to its small capacitance $C$, the QD had a charging energy $U \approx e^2/C$, a manifestation of the Coulomb repulsion between the electrons in the QD. Moreover the lateral confinement induced discrete energy levels $\varepsilon$, being broadened by the finite coupling to the leads $\Gamma$. In our QD some parameters could be easily tuned; e.g. the level positions (by means of a capacitive coupled 'plunger gate') and the tunnel barriers transparency.

If the QD is strongly decoupled from the leads ($\Gamma \leq k_B T$, with $k_B$ the Boltzmann constant and $T$ the electron temperature), transport through the QD is suppressed due to the Coulomb repulsion. Only when the charging energy is fully compensated by the plunger gate voltage, the dot's occupancy can fluctuate between $N$ and $N+1$ electrons, leading to a conductance as high as $g_{CB} = e^2/h$, $e$ being the electron charge and $h$ Plank's constant. Compared to the conductance of a single partitioned ballistic channel $g_Q = 2e^2/h$, a factor of 2 is omitted since the conductance is restricted to the occupancy of a single electron at a time, due to the electron-electron Coulomb repulsion.

A well studied many-body effect arises when the QD is strongly coupled to the leads ($\Gamma \gg k_B T$), and the dot posses a net spin, e.g. $N$ is odd. Then, the delocalized electrons in the leads arrange themselves so as to screen the net spin [4,5], forming a singlet with the impurity. This is reminiscent of the Kondo effect in metals [2], where at temperatures well below a characteristic temperature $T_K$, the Kondo temperature, the screening is complete. This temperature represents the binding energy $k_B T_K$ between the impurity and delocalized electrons in the leads. The most prominent difference between the Kondo effect in a dot and the magnetic impurity in a host metal is that it results in an *enhancement* of the conductance vs. a decrease in the conductance, respectively. Goldhaber-Gordon *et al.* [6] and Cronenwett *et al.* [7] provided conclusive evidence for the emergence of the Kondo effect in QDs. Moreover, van der Wiel *et al.* [8] showed that the enhanced conductance can reach the unitary limit, $2e^2/h$, when $T \ll T_K$. The Kondo effect had been also observed in a variety of systems, such as molecules [9, 10] and carbon nanotubes [11, 12]; with an integer spin [13], for the singlet-triplet transition [14], in the orbital form [15], and in the out-of-equilibrium regime [16, 17]. This coherent many-body state creates a peak (of width $k_B T_K$) in the density of states within the QD, pinned at the leads' Fermi level. The application of a finite bias between the leads misaligns the two peaks and the conductance is suppressed. This 'zero bias anomaly' is one of the fingerprints of the Kondo effect and is clearly seen in Fig. 1(b).

Recently, Meir et al. [18], Sela et al. [1], and Golub [19], predicted that as the QD in the unitary limit is being slightly biased and weak backscattering sets in – lowering thus the conductance - 'two-electron' backscattering processes become significant. Hence, the average 'backscattered charge' $e^*$ is larger than the electron charge. The exact value of the averaged scattered charge depends on the relative probabilities of the single- and two-electron backscattering events, turning out to be $e^* \sim 5e/3$ [1].

The magnitude of the backscattered charges cannot be obtained by measuring the transmission probability $t$ only [20]. To this end, we measured the shot noise in the current (with a low frequency spectral density $S$). For stochastic backscattering of independent charges with probability of 1-$t$, one expects in a single partitioned ballistic channel with a conductance $g=g_Q t$, a Poissonian shot noise at zero temperature [21], $S=2e^*|V_{SD}|g_Q t(1-t)$, if a bias voltage $V_{SD}$ is applied. This reduces to the well known *classical* Poissonian expression for shot noise when $t\ll 1$ (the 'Schottky equation'), $S=2e^*|I|t$. At finite bias and temperature the total noise $S_T$ is the sum of the Johnson-Nyquist noise contribution $4k_B Tg$ and the excess noise $S_{excess}$ [22]:

$$S_T = 4k_B Tg + 2e^* V_{SD} g_Q t(1-t)[coth(x)-1/x] \:, \qquad (1)$$

where $x=eV_{SD}/2k_B T$. This expression, developed for *non-interacting* charges, has been also successfully used to determine the charge of the Fractional Quantum Hall effect [23].

Referring to Fig. 1(a), the drain was connected to ground (at the cold finger) via a superconducting coil forming an *LC* resonant circuit with the cable capacitance. At resonance (of ~0.9 MHz), the *LC* circuit had a very high impedance, whereas at low frequencies was nearly zero. The differential conductance was measured at ~3 Hz with an AC current of 20-50 pA at the source superimposed to a DC current.

Shot noise in the drain was measured by a spectrum analyzer after amplification by a home-made cold amplifier [23, 24] followed by a room temperature amplifier. The measuring bandwidth was ~30 kHz. The calibration of the amplification chain was performed by measuring the thermal Johnson-Nyquist voltage noise of the sample $4k_B Tg$ versus $g=1/R$ at liquid [4]He temperature, $T=4.2$ K. Moreover, the 'cold amplifier' was characterized by a finite 'current' and 'voltage' noise, namely, $S^V_{amp}$ and $S^I_{amp}$ referred to the amplifier input. The latter was more problematic since it induced voltage fluctuations $S^I_{amp} R^2$ that were dependent on the QD resistance $R$. By referring the total noise to the input, $S^V_{amp} + 4k_B Tg + S^I_{amp} R^2$, and measuring it as a function of $R$, the amplifier current and voltage noises could be extracted, $S^V_{amp}$=1·10$^{-18}$ V$^2$/Hz, $S^I_{amp}$ =1.2·10$^{-28}$ A$^2$/Hz. The reported noise contributions are after subtraction of the thermal and amplifier noise components. Furthermore, at base temperature, the electron temperature was determined to be $T_e$~10 mK.

Shot noise was measured as a function of the transmitted current, for different couplings of the QD to the leads and at different plunger gate voltages. The results were then fitted to Eq. (1), which was modified to account for $V_{SD}$-dependent $t$, $S_{excess}$=$S(V_{SD})$ = $\sum_0^{|V|<|V_{SD}|} \frac{dS_{excess}(V,t)}{dV} \delta V$ , with the transmission $t$ being replaced by a bias dependent $t(V)$ =$g(V_{SD})/g_K$, and $\delta V$ chosen such that $t(V)$ can be regarded as constant. At a conductance of $g_{max}$~1.4$e^2/h$ the measured shot noise for a small range of applied $V_{SD}$ returned to $e^*$=(1.0±0.1)$e$, as seen in Fig. 2(a). Increasing the coupling $\Gamma$ to the leads so that $g_{max}$~1.8$e^2/h$, the shot noise increased with a nice fit to $e^*$=(1.7±0.2)$e$, as seen in Fig. 2(b).

Keeping the same coupling strength to the leads, Fig. 3 shows the noise measurements and the fits as function of the plunger gate voltage, namely, as we change $T_K$. With the plunger voltage set to the maximal conductance point of 1.8$e^2$/h, the noise was fitted with an average charge of $e^*$=(1.7±0.2)$e$. As the plunger voltage changed and the Kondo temperature lowered, so did the average charge, reaching the expected value of $e^*$=$e$.

Sela *et al.* [1] recently derived an explicit expression for the noise in the Kondo regime, close to the unitary limit, by perturbation in the small parameter $eV_{SD}/k_BT_K$<<1:

$$S_{excess} = 2eg_K |V_{SD}| \left[ \delta^2 + \frac{5}{6}\left(\frac{eV_{SD}}{k_B T_K}\right)^2 \right] \quad , \tag{2}$$

valid at zero temperature and in the limit $\delta^2 = 1 - g_{max}/g_Q \ll 1$. Experimentally, $\delta$ can be tuned by changing the barriers' asymmetry or by changing the QD levels position with the plunger voltage. To extract an effective charge from Eq. (2) and compare it to our measurements, we noted that at $t \sim 1$, $e^*$ is the ratio $S_{excess}/2I_B$, with the backscattered current $I_B = g_K V_{SD} - I_t$ expressed as a function of the transmitted current

$$I_t = g_K V_{SD} \left[ 1 - \delta^2 - \frac{1}{2}\left(\frac{eV_{SD}}{k_B T_K}\right)^2 \right] . \tag{3}$$

Eq. (2) predicts a crossover as a function of $V_{SD}$: for $eV_{SD} < \delta k_B T_K$, the first term dominates and the effective charge is $e^* = e$. In this range, namely, a rather asymmetric QD, single electron backscattering dominates. This is also the expected result in a non-interacting system, where conductance is a stochastic process of uncorrelated electron backscattering events. For $\delta k_B T_K < eV_{SD} < k_B T_K$, two-particle backscattering events also take place, leading to an increased backscattering, and an average charge of $e^* = 5e/3$. Note that the two-particle process scatters electrons with opposite spin. While the derivation of both transmitted current and shot noise assumed single level transport, this condition was *not* met in our experiment.

We then fitted in Fig. 4 the measured shot noise to the predicted one in Eq. (2). The fit used only independently measured parameters, such as $T_K \sim 30$ μeV and $\delta^2 \sim 0.3$. For example, the Kondo temperature is extracted by fitting the measured differential conductance (see inset of Fig. 4(a)) to $g(V_{SD}) = g_{max}\left[1 - \frac{3}{2}\left(\frac{eV_{SD}}{k_B T_K}\right)^2\right]$ [25]. We find a reasonable agreement up to $eV_{SD} \sim 0.5 k_B T_K = 15$ μeV, beyond which we assumed that the theoretical model was no longer valid since it assumed low applied voltage ($k_BT$, $eV_{SD}<k_BT_K$). As for $\delta^2=1-g_{max}/g_K$, the finite temperature lead to a systematic error of the order $(T/T_K)^2<1$, by lowering the maximum conductance $g_{max}$ at zero bias [25]. While Eq. (2) predicts a crossover at $eV_{SD}\sim\delta k_BT_K$, we could not resolve the two different regimes, due to the effect of the finite electron temperature and the low $V_{SD}$, resulting in a noise signal too small to be detectable. It is surprising that even when the measured noise deviates from the prediction of Eq. (2), the average backscattered charge extracted from Eq. (1) continues to indicate dominance of two-particle backscattering processes.

The experiments described here, where the backscattered charge was extracted from the spectral density of the shot noise, was not a trivial one. Since it must be a two terminal measurement, the non-linear resistance had a major effect of the spurious noise sources, which must be carefully subtracted from the total noise signal. Doing that, we indeed found, and surprisingly in a wide range of biasing voltage, what had been predicted to hold true only in a small biasing range, a backscattered average charge of $e^*\sim 5e/3$. This clearly indicates bunching of electrons as they are being partitioned by the QD in the Kondo correlated regime. Theory claims that this effect results from pairs of opposite spins being backscattered. By finding a way to separate the two-particles in the pairs, entangled separate electrons could be generated.

We thank Y. Oreg, E. Sela, F. von Oppen and Y. Meir for many helpful discussions. O. Z. acknowledges support from the Israeli Ministry of Science and Technology. This work was partly supported by the MINERVA Foundation, the German Israeli Foundation (GIF), the German Israeli Project cooperation (DIP) and the Israeli Science Foundation (ISF).

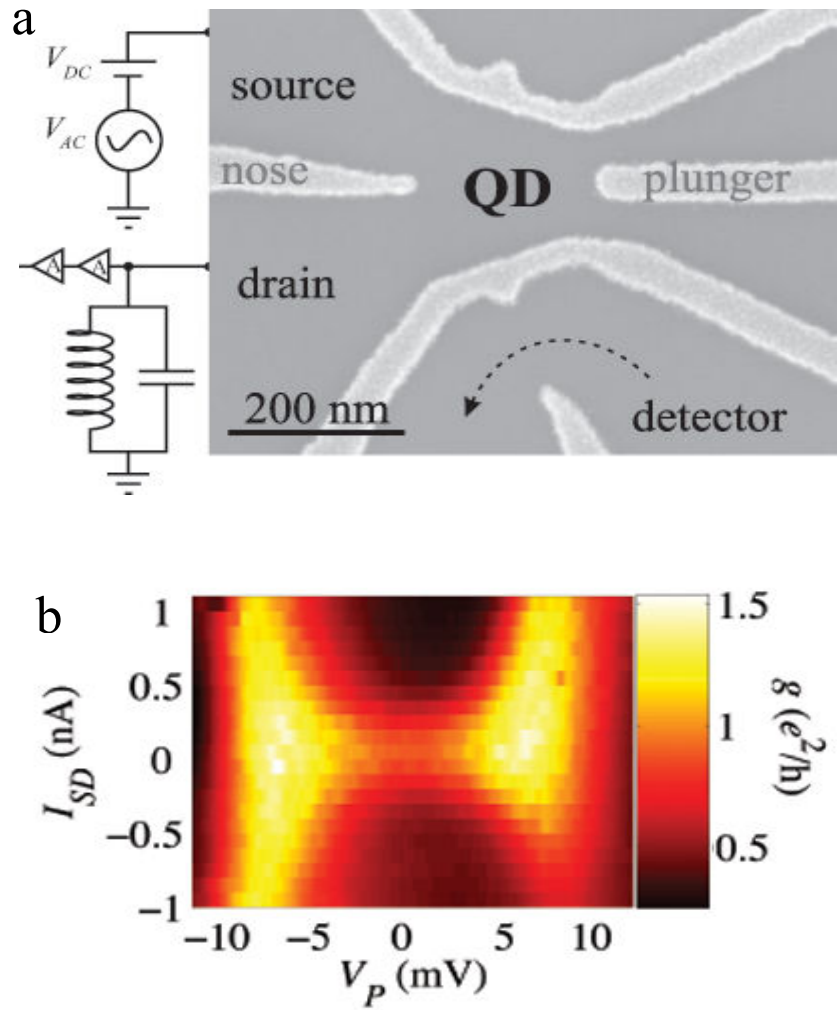

Figure 1: **Measurement scheme and Kondo's zero bias anomaly.** (a) SEM micrograph of the device embedded in a GaAs-AlGaAs heterostructure, supporting a 2DEG with density $3.1 \cdot 10^{11}$ cm$^{-2}$ and mobility $2.3 \cdot 10^{6}$ cm$^{2}$/Vs at 4.2 K. The QD was formed by biasing the metallic gates patterned by e-beam lithography. The conductance was measured by forcing an AC current superimposed to a DC current though the source and measuring the voltage with lock-in technique. An inductor was placed in series at the QD drain, to form a resonant circuit with the cable capacitance at ~0.9 MHz, followed by a home-made cold amplifier and a room temperature amplifier. (b) Conductance of a Kondo resonance versus plunger voltage, $V_P$, and source drain current, $I_{SD}$.

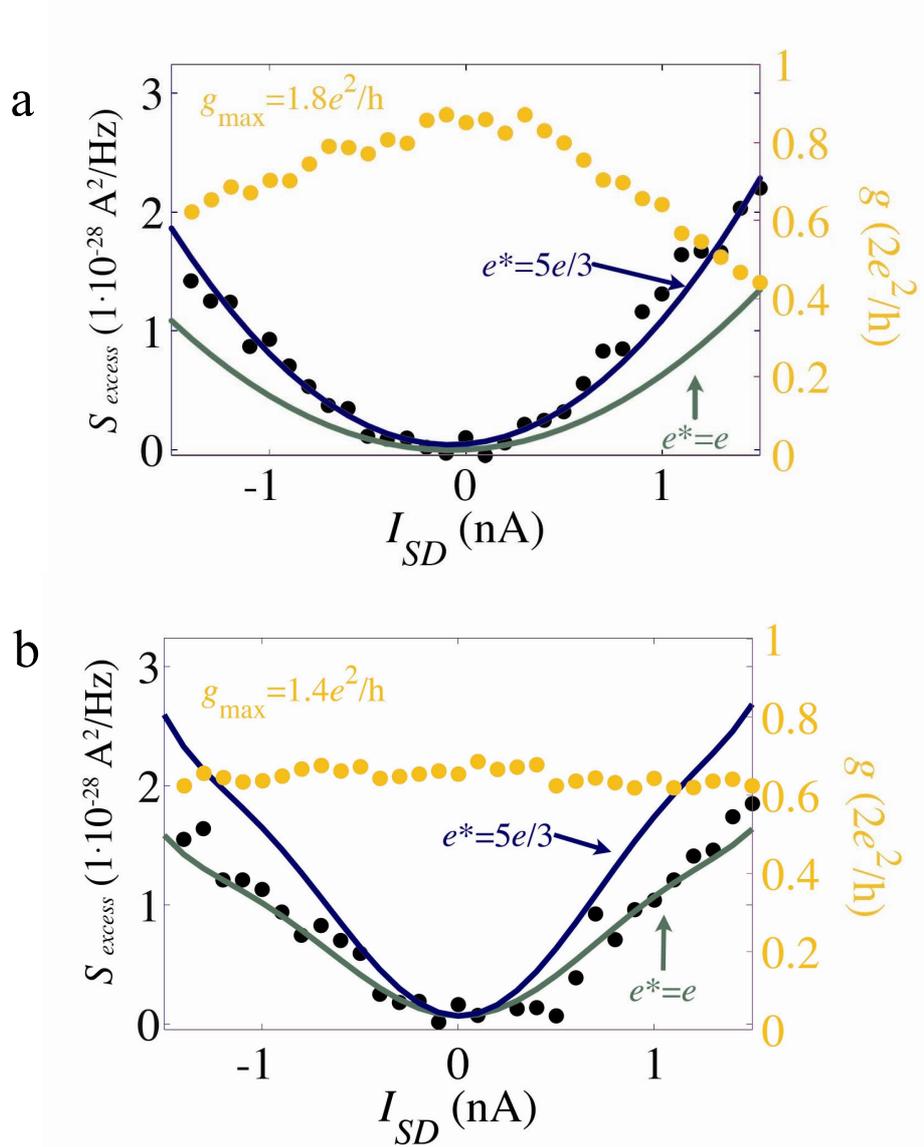

Figure 2: **Shot noise near the unitary limit.** (a, b) Excess shot noise, $S_{excess}$ (black circles), at a maximal conductance of $g_{max}=1.8e^2/h$ and $g_{max}=1.4e^2/h$, fitted with $e^*=5e/3$ (blue) and $e^*=e$ (green). Right axes display measured conductance (orange).

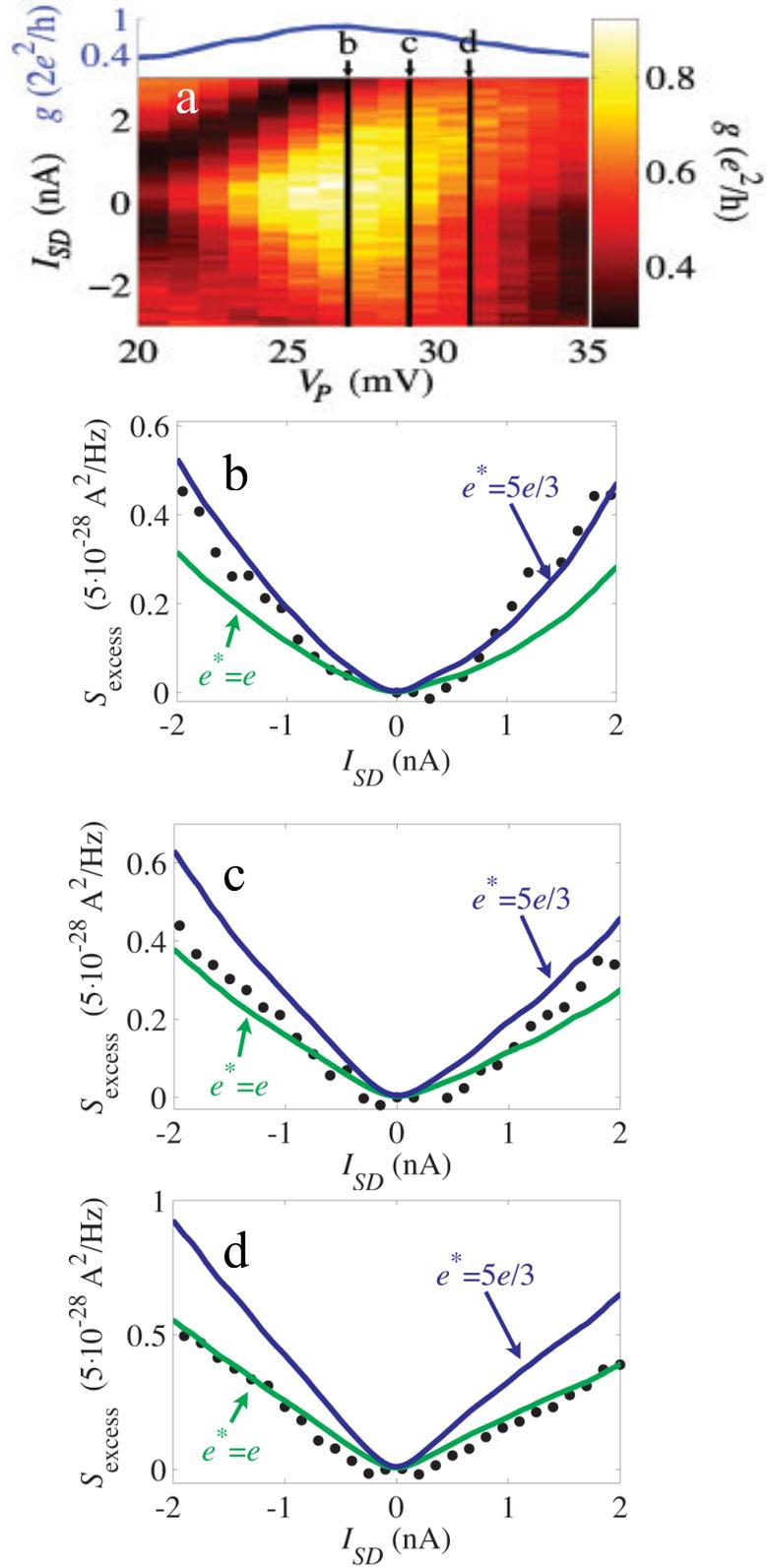

Figure 3: **Evolution of shot noise along the Kondo enhanced conductance peak.** (a) Conductance versus plunger voltage $V_P$, and source-drain current $I_{SD}$. The trace above is a cut through $I_{SD}=0$. (b-d) Plots of shot noise at and away from the conductance peak, with a theoretical fit to $e^*=5e/3$ (blue) and $e^*=e$ (green).

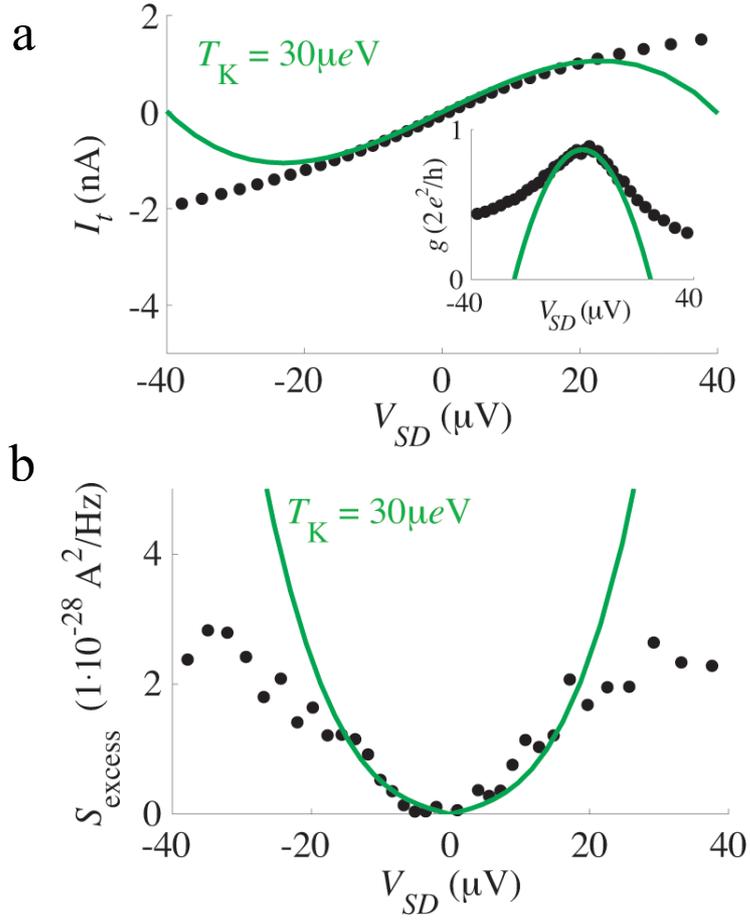

Figure 4: **Comparison with theory of the conductance, transmitted current and noise.** (a) Comparison between the measured transmitted current $I_t$ (black circles) and calculated one (green line). Inset: Differential conductance versus $V_{SD}$ (black circles) and theoretical best fit (green line) to the conductance from which a Kondo temperature $T_K$ =30µeV was extracted. (b) Comparison between the measured excess noise, $S_{excess}$ (black circles) and the prediction of Eq. (2) (green line).